\documentclass[aps,prc,preprint,showpacs,showkeys]{revtex4}
\usepackage{amssymb}
\usepackage{amsmath}
\usepackage{graphicx}
\usepackage{bm}
\usepackage{epsf}

\begin{document}

\title{Properties of hyperonic matter in strong magnetic fields}
\author{P. Yue}
\affiliation{Department of Physics, Nankai University, Tianjin 300071, China}
\author{F. Yang}
\affiliation{Department of Physics, Nankai University, Tianjin 300071, China}
\author{H. Shen}
\email{songtc@nankai.edu.cn}
\affiliation{Department of Physics, Nankai University, Tianjin 300071, China}

\begin{abstract}
We study the effects of strong magnetic fields on the properties of hyperonic matter.
We employ the relativistic mean field theory,
which is known to provide excellent descriptions of nuclear matter and finite
nuclei. The two additional hidden-strangeness mesons, $\sigma^{\ast}$ and $\phi$,
are taken into account, and some reasonable hyperon potentials are used
to constrain the meson-hyperon couplings, which reflect the recent
developments in hypernuclear physics. It is found that the effects
of strong magnetic fields become significant only for magnetic field
strength $B>5 \times 10^{18}$ G. The threshold densities of hyperons
can be significantly altered by strong magnetic fields. The presence of hyperons
makes the equation of state (EOS) softer than that in the case without hyperons,
and the softening of the EOS becomes less pronounced with increasing magnetic
field strength.
\end{abstract}


\maketitle


\section{Introduction}

\label{sec:1}

Neutron stars are fascinating celestial objects as natural laboratories for dense
matter physics. They are associated with the most exotic environments and violent
phenomena of the universe~\cite{Weber05}.
Observations of ordinary radio pulsars indicate that they possess surface magnetic
fields of the order of $10^{12}$ G~\cite{HL06}. Recent surveys of soft-gamma
repeaters (SGRs) and anomalous x-ray pulsars (AXPs) imply that the surface magnetic
field of young neutron stars could be of order $10^{14}$--$10^{15}$ G~\cite{TD96}.
The magnetic field strength may vary significantly from the surface to the center
in neutron stars. So far, there is no direct observational evidence for the internal
magnetic fields of the star, while it may reach $10^{18}$ G, as estimated in
some theoretical works~\cite{Latt00,Latt01,Mao03}.
There are many studies of the effects of strong magnetic fields on the properties of
dense nuclear matter and neutron
stars~\cite{HL06,Latt00,Latt01,Mao03,CBP97,Mao06,YS06},
and the inclusion of hyperons and boson condensation has also been
investigated in the literature~\cite{Latt02,Pion01,Pion07,Kaon02,YS08}.
On the other hand, much work has been done to investigate the effects of the magnetic
field on color superconducting phases of dense quark matter, which are conjectured
to exist inside neutron stars~\cite{CSCPRL05NPB06,CSCPRL06PRD07,CSCPRD07,CSCPRL08}.

Our knowledge of neutron star interiors is still uncertain.
The central density of neutron stars can be extremely high,
and many possibilities for such dense matter have
been suggested~\cite{Weber05,PR00}.
For densities below twice normal nuclear matter density
($\rho_{0} \sim 0.15\;\mathrm{fm}^{-3}$), the matter consists of only nucleons
and leptons. When the density is higher than $2 \rho_{0}$, the equation
of state (EOS) and composition of matter are much less certain.
The presence of hyperons in neutron stars has been studied by many
authors~\cite{PRC96,PRC99,Panda04,Shen02,Gl01,Shen03}.
With increasing density, hyperons may appear through the weak interaction
due to the fast rise of the baryon chemical potentials.
The presence of hyperons tends to soften the EOS at high density
and lower the maximum mass of neutron
stars~\cite{PR00,PRC96,PRC99,Panda04,Shen02,Gl01,Shen03}.
$K^{-}$ condensation in dense matter was suggested by Kaplan and
Nelson~\cite{KN86} and has been discussed in many
works~\cite{PR00,PRC96,Kaon02,YS08}. In general, the chemical
potential of antikaons decreases with increasing density because of
the interaction with baryons. As a consequence, the ground state of
matter at high density may contain antikaon condensation which can
soften the EOS of neutron star matter.
It has been suggested that the quark matter may exist in the core of
massive neutron stars, and the hadron-quark phase transition can
proceed through a mixed phase of hadronic and quark
matter~\cite{Weber05,PR00,Panda04,PRC08}. If deconfined quark matter
does exist inside stars, it is likely to be in a color superconducting
phase~\cite{Weber05,Panda04}. At present, the densities at which
these phases occur are still rather uncertain.

In this article, we focus on the properties of hyperonic matter,
which is composed of a chemically equilibrated and charge-neutral
mixture of nucleons, hyperons, and leptons.
It is well known that hyperons may appear around twice normal nuclear
matter density and can soften the EOS at high density
in the field-free case~\cite{PR00,PRC96,PRC99,Panda04,Shen02}.
The meson-hyperon couplings play an important role in determining
the EOS and composition of hyperonic matter~\cite{Gl01,Shen03}.
In the presence of strong magnetic fields, the pressure and composition of
matter can be affected significantly. In Ref.~\cite{Latt02}, the authors
investigated the effects of strong magnetic fields on the properties
of neutron star matter including hyperons. They employed two classes
of the relativistic mean field (RMF) models, GM1-3 and ZM models,
while the meson-hyperon couplings were assumed to be equal for all hyperons
in their study. It was found that the EOS at high density could be significantly
affected both by Landau quantization and by magnetic moment interactions,
but only for field strength $B > 5 \times 10^{18}$ G.
It is very interesting to investigate the influence of strong magnetic fields
on hyperonic matter with recent developments in hypernuclear physics.
A recent observation of the double-$\Lambda$ hypernucleus
$_{\Lambda\Lambda}^{6}\mathrm{He}$, called the Nagara event~\cite{Nagara},
has had a significant impact on strangeness nuclear physics. The Nagara event
provides unambiguous identification of $_{\Lambda\Lambda}^{6}\mathrm{He}$
production with a precise $\Lambda\Lambda$ binding energy value
$B_{\Lambda\Lambda }=7.25\pm 0.19_{-0.11}^{+0.18}\;\mathrm{MeV}$,
which suggests that the effective $\Lambda\Lambda$ interaction should be
considerably weaker ($\triangle B_{\Lambda\Lambda}\simeq 1\;\mathrm{MeV}$)
than that deduced from the earlier measurement
($\triangle B_{\Lambda\Lambda }\simeq$ 4--5 MeV).
The weak hyperon-hyperon ($YY$) interaction suggested by the Nagara event
has been used to reinvestigate the properties of multistrange systems,
and it has been found that the change of $YY$ interactions affects the
properties of strange hadronic matter dramatically~\cite{PRC03s,JPG04s,JPG05}.
We would like to examine how the $YY$ interaction affects the properties of
hyperonic matter in the presence of strong magnetic fields.

We employ the RMF theory to study the properties of hyperonic matter
in the presence of strong magnetic fields. The RMF theory has been
successfully and widely used for the description of nuclear matter and finite
nuclei~\cite{Serot86,Ring90,Toki96,Ren02,Shen06}. It has also been applied to
providing the EOS of dense matter for use in supernovae and neutron
stars~\cite{Shen98}. In the RMF approach, baryons interact through the exchange
of scalar and vector mesons. The meson-nucleon coupling constants are generally
determined by fitting to some nuclear matter properties or ground-state properties of
finite nuclei. However, there are large uncertainties in the meson-hyperon couplings
due to limited experimental data in hypernuclear physics. Generally, one can use the
coupling constants derived from the quark model or the values constrained by
reasonable hyperon potentials. The two additional hidden-strangeness mesons,
$\sigma^{\ast}$ and $\phi$, were originally introduced to obtain the
strong attractive $YY$ interaction deduced from the earlier measurement~\cite{PRL93}.
Their couplings to hyperons are related to the $YY$ interaction~\cite{PRC08}.
In this study, we adopt the TM1 version of the RMF theory, which includes
nonlinear terms for both $\sigma$ and $\omega$ mesons~\cite{ST94}.
The TM1 model is known to provide excellent descriptions of the ground
states of finite nuclei including unstable nuclei and also shown to reproduce
satisfactory agreement with experimental data in the studies of nuclei
with deformed configuration and the giant resonances within the
random-phase approximation (RPA) formalism~\cite{HTT95,MT97,MV01}.
As for the meson-hyperon couplings,
we use the values constrained by reasonable hyperon potentials that include
the updated information from recent developments in hypernuclear physics.
We take into account the two additional hidden-strangeness mesons, $\sigma^{\ast}$
and $\phi$, and consider two cases of the $YY$ interaction, the weak and
strong $YY$ interactions. By comparing the results with different $YY$
interactions, we evaluate how sensitive the results are to this factor.
Because the properties of hyperonic matter are very sensitive to the
meson-hyperon couplings, an investigation of the influence of strong
magnetic fields on hyperonic matter with a suitable choice of the meson-hyperon
couplings would be of interest for the study of neutron stars.

This article is arranged as follows. In Sec.~\ref{sec:2}, we briefly describe
the RMF theory for hyperonic matter in the presence of
strong magnetic fields. In Sec.~\ref{sec:3}, we discuss the parameters used
in the calculation. We show and discuss the numerical results in Sec.~\ref{sec:4}.
Section~\ref{sec:5} is devoted to a summary.


\section{ Formalism}

\label{sec:2}

We briefly explain the RMF theory used to describe hyperonic matter
in the presence of strong magnetic fields. In the RMF approach, baryons
interact through the exchange of scalar and vector mesons.
The baryons considered in this work are nucleons ($p$ and $n$) and
hyperons ($\Lambda$, $\Sigma$, and $\Xi$). The exchanged mesons include
isoscalar scalar and vector mesons ($\sigma$ and $\omega$), isovector vector
meson ($\rho$), and two additional hidden-strangeness mesons
($\sigma^{\ast }$ and $\phi$).
In the presence of strong magnetic fields, the total Lagrangian density of
matter consisting of a neutral mixture of baryons and leptons
takes the form
\begin{eqnarray}
\mathcal{L}_{\mathrm{RMF}} &=& \sum_{b}\bar{\psi}_{b}
\left[ i\gamma_{\mu}\partial^{\mu}
-m_{b}-g_{\sigma b}\sigma-g_{\sigma^{\ast} b}\sigma^{\ast} \right. \nonumber\\
&& -g_{\omega b}\gamma_{\mu}\omega^{\mu}-g_{\phi b}\gamma_{\mu}\phi^{\mu}
-g_{\rho b}\gamma_{\mu}\tau_{ib}\rho_{i}^{\mu} \nonumber\\
&&\left. -q_{b}\gamma_{\mu}A^{\mu}
-\frac{1}{2}\kappa_{b}\sigma_{\mu\nu}F^{\mu\nu}\right]
\psi_{b} \nonumber \\
&&+\sum_{l}\bar{\psi}_{l}\left[ i\gamma_{\mu}\partial^{\mu}-m_{l}
-q_{l}\gamma_{\mu}A^{\mu} \right]
\psi_{l} \nonumber \\
&& +\frac{1}{2}\partial_{\mu}\sigma\partial^{\mu}\sigma
   -\frac{1}{2} m_{\sigma}^{2}\sigma^{2}-\frac{1}{3}g_{2}\sigma^{3}
   -\frac{1}{4}g_{3}\sigma^{4} \nonumber \\
&& -\frac{1}{4}W_{\mu\nu}W^{\mu\nu}+\frac{1}{2}m_{\omega}^{2}\omega_{\mu}\omega^{\mu}
   +\frac{1}{4}c_{3}\left(\omega_{\mu}\omega^{\mu}\right)^{2} \nonumber \\
&& -\frac{1}{4}R_{i\mu\nu}R_{i}^{\mu\nu}
   +\frac{1}{2}m_{\rho}^{2}\rho_{i\mu}\rho_{i}^{\mu}
   +\frac{1}{2}\partial_{\mu}\sigma^{\ast}\partial^{\mu}\sigma^{\ast}  \nonumber \\
&& -\frac{1}{2}m_{\sigma^{\ast}}^{2}\sigma^{\ast2}-\frac{1}{4}S_{\mu\nu}S^{\mu\nu}
   +\frac{1}{2}m_{\phi}^{2}\phi_{\mu}\phi^{\mu}-\frac{1}{4}F_{\mu\nu}F^{\mu\nu},
\label{eq:lrmf}
\end{eqnarray}
where $\psi_{b}$ and $\psi_{l}$ are the baryon and lepton fields,
respectively. The index $b$ runs over the baryon octet ($p$, $n$,
$\Lambda$, $\Sigma^{+}$, $\Sigma^{0}$, $\Sigma^{-}$, $\Xi^{0}$,
$\Xi^{-}$), and the sum on $l$ is over electrons and muons
($e^{-}$ and $\mu^{-}$). The field tensors of the vector mesons,
$\omega$, $\rho$, and $\phi$, are denoted by $W_{\mu\nu}$,
$R_{i\mu\nu}$, and $S_{\mu\nu}$, respectively.
$A^{\mu}=(0$, $0$, $Bx$, $0)$ refers to a constant external magnetic
field $B$ along the $z$ axis with its field tensor denoted by $F_{\mu\nu}$.
It has been found that the contributions from the anomalous magnetic moments
of baryons should be considered in strong magnetic
fields~\cite{Latt00,Latt01,Mao03,Mao06,YS06}. The anomalous
magnetic moments are introduced via the minimal coupling of the
baryons to the electromagnetic field tensor in the term
$-\frac{1}{2}\kappa_{b}\sigma_{\mu\nu}F^{\mu\nu}$, where
$\kappa_b = (\mu_b/\mu_N - q_b m_p/m_b)~\mu_N$ is the anomalous magnetic
moment of baryon species $b$ ($\mu_N=e\hbar/2m_p$ is the nuclear magneton).
As for leptons, the effects of anomalous magnetic moments
on the EOS are very small, as shown in Ref.~\cite{Mao06},
and the electron anomalous magnetic moments could be efficiently reduced
by high-order contributions from the vacuum polarization in strong magnetic
fields~\cite{eAMM}. Therefore, we neglect the anomalous magnetic moments of
leptons in the present work.

In the RMF approach, the meson fields are treated as classical fields,
and the field operators are replaced by their expectation values.
The meson field equations in uniform matter have the following form:
\begin{eqnarray}
&& m_{\sigma }^{2}\sigma +g_{2}\sigma ^{2}+g_{3}\sigma ^{3}
=-\sum_{b} g_{\sigma b}\rho_{s}^{b}, \label{eq:s} \\
&& m_{\omega }^{2}\omega +c_{3}\omega ^{3} =\sum_{b}g_{\omega b}\rho_{v}^{b},
\label{eq:w} \\
&& m_{\rho }^{2}\rho =\sum_{b}g_{\rho b}\tau_{3b}\rho_{v}^{b},  \label{eq:r}\\
&& m_{\sigma^{\ast}}^{2}\sigma^{\ast}=-\sum_{b}g_{\sigma^{\ast}b}\rho_{s}^{b},
\label{eq:ss}\\
&& m_{\phi}^{2}\phi=\sum_{b}g_{\phi b}\rho_{v}^{b},  \label{eq:ws}
\end{eqnarray}
where $\sigma =\left\langle \sigma \right\rangle$,
$\omega =\left\langle \omega^{0}\right\rangle$,
$\rho =\left\langle \rho_{3}^{0}\right\rangle$,
$\sigma^{\ast }=\left\langle \sigma ^{\ast }\right\rangle$, and
$\phi =\left\langle \phi ^{0}\right\rangle$ are the nonvanishing expectation
values of meson fields in uniform matter.
In the presence of strong magnetic fields, the Dirac equations for baryons
and leptons are, respectively, given by
\begin{eqnarray}
&& \left( i\gamma_{\mu}\partial^{\mu}-m_{b}^{\ast}
-g_{\omega b}\gamma_{0}\omega-g_{\phi b}\gamma_{0}\phi
-g_{\rho b}\gamma_{0}\tau_{3b}\rho \right.\nonumber\\
&& \left. -q_{b}\gamma_{\mu}A^{\mu}
-\frac{1}{2}\kappa_{b}\sigma_{\mu\nu}F^{\mu\nu}\right)\psi_{b} = 0, \\
&& \left( i\gamma_{\mu}\partial^{\mu }-m_{l}
-q_{l}\gamma_{\mu }A^{\mu}\right)\psi_{l} = 0,
\end{eqnarray}
where $m_{b}^{\ast}=m_{b}+g_{\sigma b}\sigma +g_{\sigma^{\ast} b}\sigma^{\ast}$
is the effective mass of baryon species $b$.
Following the method of Refs.~\cite{Latt00,Mao03,PRD71},
we solve the Dirac equations and obtain the energy spectra
for charged and uncharged baryons as
\begin{eqnarray}
E_{\nu,s}^{b} &=& \sqrt{k_{z}^{2}+\left( \sqrt{ m_{b}^{\ast 2}+2\nu |q_{b}| B}
-s{\kappa_{b}B}\right)^{2}} +g_{\omega b}\omega +g_{\phi b}\phi
+g_{\rho b}\tau_{3b}\rho, \\
E_{s}^{b} &=& \sqrt{k_{z}^{2}+\left( \sqrt{m_{b}^{\ast 2}+k_{x}^{2}+k_{y}^{2}}
-s{\kappa_{b}B}\right)^{2}} +g_{\omega b}\omega
+g_{\phi b}\phi +g_{\rho b}\tau_{3b}\rho,
\end{eqnarray}
while the lepton energy spectrum is expressed by
\begin{eqnarray}
E_{\nu,s}^{l} &=& \sqrt{k_{z}^{2}+m_{l}^{2}{+}2\nu \left\vert {q_{l}}%
\right\vert {B}}.
\end{eqnarray}
Here $\nu=0,1,2,\ldots$ enumerates the Landau levels of charged fermions.
The quantum number $s$ is equal to $\pm 1$. For charged baryons,
the scalar and vector densities are, respectively, given by~\cite{Latt00,Mao03}
\begin{eqnarray}
\rho_{s}^{b} &=& \frac{{|q_{b}|B}m_{b}^{\ast }}{2\pi^{2}}\sum_{\nu}\sum_{s}
\frac{\sqrt{m_{b}^{\ast 2}+2\nu {|q_{b}|B}}-s\kappa_{b}B}
{\sqrt{m_{b}^{\ast 2}+2\nu {|q_{b}|B}}}
\ln \left\vert \frac{k_{f,\nu ,s}^{b}+E_{f}^{b}}
{\sqrt{m_{b}^{\ast 2}+2\nu {|q_{b}|B}}-s\kappa_{b}B}\right\vert  , \\
\rho_{v}^{b} &=& \frac{{|q_{b}|}B}{2\pi ^{2}}\sum_{\nu}\sum_{s}k_{f,\nu,s}^{b},
\end{eqnarray}
where $k_{f,\nu ,s}^{b}$  is the Fermi momentum of charged baryon $b$ with
quantum numbers $\nu$ and $s$.  The Fermi energy $E_{f}^{b}$ is related to the Fermi
momentum $k_{f,\nu ,s}^{b}$ by
\begin{eqnarray}
{E_{f}^{b}}^2 &=& \left(k_{f,\nu ,s}^{b}\right)^2
+\left( \sqrt{m_{b}^{\ast 2}+2\nu {\vert q_{b}\vert B}}
-s\kappa_{b}B\right)^{2}.
\end{eqnarray}
For uncharged baryons, there is no Landau level quantum number $\nu$,
so the Fermi momentum is denoted by $k_{f,s}^{b}$.
The Fermi energy $E_{f}^{b}$ is related to the Fermi momentum $k_{f,s}^{b}$ by
\begin{eqnarray}
{E_{f}^{b}}^2 &=& \left(k_{f,s}^{b}\right)^2
+\left( m_{b}^{\ast }-s\kappa_{b}B\right)^{2}.
\end{eqnarray}
The scalar and vector densities of uncharged baryon $b$ are,
respectively, given by~\cite{Latt00,Mao03}
\begin{eqnarray}
\rho_{s}^{b} &=& \frac{m_{b}^{\ast}}{4\pi ^{2}}\sum_{s}
\left[ k_{f,s}^{b}E_{f}^{b}-\left( m_{b}^{\ast }-s{\kappa_{b}B}\right)^{2}
\ln \left\vert \frac{k_{f,s}^{b}+E_{f}^{b}}{m_{b}^{\ast }
-s{\kappa_{b}B}}\right\vert \right], \\
\rho_{v}^{b} &=& \frac{1}{2\pi^{2}}\sum_{s}
\left\{ \frac{1}{3} \left(k_{f,s}^{b}\right)^3
-\frac{1}{2}s\kappa_{b}B\left[ \left( m_{b}^{\ast}-s{\kappa_{b}B}\right)k_{f,s}^{b}
\right.\right.\nonumber\\
& &\left.\left. +{E_{f}^{b}}^2 \left( \arcsin \frac{m_{b}^{\ast}-s{\kappa_{b}B}}
{E_{f}^{b}}-\frac{\pi}{2}\right) \right] \right\} .
\end{eqnarray}

For hyperonic matter consisting of a neutral mixture of baryons and
leptons, the $\beta$-equilibrium conditions without trapped neutrinos are
given by
\begin{eqnarray}
&&\mu_{p}=\mu_{\Sigma^{+}}=\mu_{n}-\mu_{e},\label{eq:beta1}\\
&&\mu_{\Lambda}=\mu_{\Sigma^{0}}=\mu_{\Xi^{0}}=\mu_{n},\label{eq:beta2}\\
&&\mu_{\Sigma^{-}}=\mu_{\Xi^{-}}=\mu_{n}+\mu_{e},\label{eq:beta3}\\
&&\mu_{\mu}=\mu_{e},\label{eq:beta4}
\end{eqnarray}
where $\mu_{i}$ is the chemical potential of species $i$.
The chemical potentials of baryons and leptons are, respectively, given by
\begin{eqnarray}
\mu_{b} &=& E_{f}^{b}+g_{\omega b}\omega+g_{\phi b}\phi
                     +g_{\rho b}\tau_{3b}\rho,\label{eq:mub}\\
\mu_{l} &=& E_{f}^{l} = \sqrt{ \left(k_{f,\nu,s}^{l}\right)^2 + m_{l}^{2}
                        + 2\nu\left\vert{q_{l}}\right\vert B}.
\end{eqnarray}
The electric charge neutrality condition is expressed by
\begin{eqnarray}
\sum_b q_{b}\rho_v^{b} + \sum_l q_{l}\rho_v^{l} = 0 ,
\label{eq:charge}
\end{eqnarray}
where the vector density of leptons has a similar expression to that of
charged baryons
\begin{equation}
\rho_{v}^{l}=\frac{\left\vert {q_{l}}\right\vert B}{2\pi^{2}}
\sum_{\nu }\sum_{s}k_{f,\nu ,s}^{l}.
\end{equation}
We solve the coupled Eqs.~(\ref{eq:s})--(\ref{eq:ws}),
(\ref{eq:beta1})--(\ref{eq:beta4}), and (\ref{eq:charge}) self-consistently
at a given baryon density
$\rho_{b}=\sum_{b} \rho_{v}^{b}$ in the presence of strong magnetic fields.
The total energy density of matter is given by
\begin{eqnarray}
\varepsilon &=& \sum_b\varepsilon_{b}+\sum_l\varepsilon_{l}
+\frac{1}{2}m_{\sigma }^{2}\sigma ^{2}+\frac{1 }{3}g_{2}\sigma^{3}
+\frac{1}{4}g_{3}\sigma^{4}+\frac{1}{2}m_{\omega }^{2}\omega^{2}\nonumber\\
&&+\frac{3}{4}c_{3}\omega^{4}+\frac{1}{2}m_{\rho }^{2}\rho^{2}
+\frac{1}{2}m_{\sigma^{\ast}}^{2}\sigma^{\ast 2}+\frac{1}{2}m_{\phi}^{2}\phi^{2},
\label{eq:e1}
\end{eqnarray}
where the energy densities of charged baryons and leptons have the following forms:
\begin{eqnarray}
\varepsilon_{b} &=& \frac{|{q_{b}|B}}{4\pi ^{2}}\sum_{\nu}\sum_{s}
\left[ k_{f,\nu ,s}^{b}E_{f}^{b}+\left( \sqrt{m_{b}^{\ast 2}+2\nu |{q_{b}|B}}
-s\kappa_{b}B\right)^{2}\right.  \nonumber\\
&&\times \left. \ln \left\vert \frac{k_{f,\nu ,s}^{b}
+E_{f}^{b}}{\sqrt{m_{b}^{\ast 2}+2\nu {|q_{b}|B}}
-s\kappa _{b}B}\right\vert \right] , \\
\varepsilon_{l} &=& \frac{\left\vert {q_{l}}\right\vert B}{4\pi ^{2}}
\sum_{\nu }\sum_{s}\left[ k_{f,\nu ,s}^{l}E_{f}^{l}+\left( m_{l}^{2}
+2\nu \left\vert {q_{l}}\right\vert {B}\right) \ln \left\vert \frac{k_{f,\nu,s}^{l}
+E_{f}^{l}}{\sqrt{m_{l}^{2}{+}2\nu\left\vert{q_{l}}\right\vert{B}}}
\right\vert \right],
\end{eqnarray}
while those of uncharged baryons are given by
\begin{eqnarray}
\varepsilon_{b} &=& \frac{1}{4\pi ^{2}}\sum_{s}\left\{ \frac{1}{2}
k_{f,s}^{b}{E_{f}^{b}}^3-\frac{2}{3}s\kappa _{b}B {E_{f}^{b}}^3 \left( \arcsin
\frac{m_{b}^{\ast }-s{\kappa _{b}B}}{E_{f}^{b}}-\frac{\pi }{2}\right)
-\left( \frac{s{\kappa _{b}B}}{3}+\frac{m_{b}^{\ast }-s{\kappa_{b}B}}{4}\right)
\right.  \nonumber\\
&&\times \left. \left[ \left( m_{b}^{\ast }-s{\kappa _{b}B}\right)
k_{f,s}^{b}E_{f}^{b}+\left( m_{b}^{\ast }-s{\kappa _{b}B}\right)^{3}
\ln \left\vert \frac{k_{f,s}^{b}+E_{f}^{b}}{m_{b}^{\ast }-s{\kappa _{b}B}}
\right\vert \right] \right\} .
\end{eqnarray}
The pressure of the system can be obtained by
\begin{equation}
P=\sum_{b}\mu_{b}\rho_{v}^{b}+\sum_{l}\mu_{l}\rho_{v}^{l}-\varepsilon
 =\mu_{n}\rho_{b}-\varepsilon,
\end{equation}
where the electric charge neutrality and $\beta $-equilibrium conditions are used to
get the last equality. We note that the contribution from electromagnetic fields
to the energy density and pressure, $\varepsilon_{f}=P_{f}=B^{2}/8\pi$, is not
taken into account in the present calculation. In general, the strong magnetic
fields in neutron stars can produce magnetic forces that play an important role
in determining the structure of the star~\cite{Latt01}.


\section{ Parameters}

\label{sec:3}

In this section, we discuss the parameters used in the RMF approach.
For the nucleonic sector, we employ the TM1 parameter set which
was determined by a least-squares fit to experimental results including
stable and unstable nuclei~\cite{ST94}. It has been shown that the RMF theory
with the TM1 parameter set reproduces satisfactory agreement with experimental
data in the studies of the nuclei with deformed configuration and the giant
resonances within the RPA formalism~\cite{HTT95,MT97,MV01}.
With the TM1 parameter set, the nuclear matter saturation
density is $0.145$ fm$^{-3}$, the energy per nucleon is $-16.3$ MeV,
the symmetry energy is $36.9$ MeV,
and the incompressibility is $281$ MeV~\cite{ST94}.
The TM1 parameter set can be found in Refs.~\cite{PRC08,ST94}.

The baryons considered in this work are the baryon octet ($p$, $n$,
$\Lambda$, $\Sigma^{+}$, $\Sigma^{0}$, $\Sigma^{-}$, $\Xi^{0}$, $\Xi^{-}$).
In Table~\ref{tab:1}, we list the static properties of baryons such as
masses, charges, and magnetic moments. The exchanged mesons include isoscalar
scalar and vector mesons ($\sigma$ and $\omega$), isovector vector meson ($\rho$),
and two additional hidden-strangeness mesons ($\sigma^{\ast}$ and $\phi$).
As for the meson-hyperon couplings, we take the naive quark model values for
the vector coupling constants,
\begin{eqnarray}
&&\frac{1}{3}g_{\omega N}=\frac{1}{2}g_{\omega \Lambda }
=\frac{1}{2}g_{\omega \Sigma}=g_{\omega \Xi },  \nonumber \\
&& g_{\rho N}=\frac{1}{2}g_{\rho \Sigma }=g_{\rho \Xi },\ \ g_{\rho \Lambda}=0,
\nonumber \\
&& 2g_{\phi \Lambda}=2g_{\phi \Sigma}=g_{\phi \Xi}
  =-\frac{2\sqrt{2}}{3}g_{\omega N},\ \ g_{\phi N}=0.
\end{eqnarray}
The scalar coupling constants are chosen to give reasonable hyperon
potentials. We denote the potential depth of the hyperon species $i$ in the
matter of the baryon species $j$ by $U_{i}^{\left( j\right) }$.
It is estimated from the experimental data of single-$\Lambda$ hypernuclei that
the potential depth of a $\Lambda$ in saturated nuclear matter should be around
$U_{\Lambda}^{\left(N\right)} \simeq -30$ MeV~\cite{PRC00}. For $\Sigma$ hyperons,
the analysis of $\Sigma$ atomic experimental data suggests that $\Sigma$-nucleus
potentials have a repulsion inside the nuclear surface and an attraction outside
the nucleus with a sizable absorption. In recent theoretical works,
the $\Sigma$ potential in saturated nuclear matter is considered to be repulsive
with the strength of about $30$ MeV~\cite{PRC00,JPG08}.
Recent developments in hypernuclear physics suggest that $\Xi$ hyperons in saturated
nuclear matter have an attraction of around $15$ MeV~\cite{JPG08,PRC00xi}.
In this article, we use
$U_{\Lambda }^{\left(N\right) }=-30$,
$U_{\Sigma  }^{\left(N\right) }=+30$, and
$U_{\Xi     }^{\left(N\right) }=-15$ MeV
to determine the scalar coupling constants
$g_{\sigma\Lambda}$, $g_{\sigma\Sigma}$, and $g_{\sigma\Xi}$, respectively.

The hyperon couplings to the hidden-strangeness meson $\sigma^{\ast}$ are
restricted by the relation
$U_{\Xi }^{\left(\Xi\right) }\simeq U_{\Lambda}^{\left(\Xi\right) }
\simeq 2U_{\Xi}^{\left(\Lambda\right) }
\simeq 2U_{\Lambda }^{\left(\Lambda\right) }$ obtained in Ref.~\cite{ANN94}.
We consider two cases of hyperon-hyperon ($YY$) interactions.
The weak $YY$ interaction implied by the Nagara event suggests
$U_{\Lambda}^{(\Lambda )}\simeq -5$ MeV, while the strong $YY$ interaction
deduced from the earlier measurement suggests
$U_{\Lambda }^{(\Lambda )}\simeq -20$ MeV~\cite{PRC08,PRC03s,JPG04s,JPG05}.
In Table~\ref{tab:2}, we present the meson-hyperon couplings determined
by these hyperon potentials. We note that $m_{\sigma^{\ast }}=980$ MeV
and $m_{\phi }=1020$ MeV are used in the present work.


\section{ Results and discussion}

\label{sec:4}

In this section, we investigate the properties of hyperonic matter in the
presence of strong magnetic fields using the RMF theory. We adopt the TM1 version
of the RMF theory, which is known to provide excellent descriptions of the ground
states of finite nuclei including unstable nuclei~\cite{ST94}.
For the meson-hyperon couplings, we use the values constrained by reasonable
hyperon potentials that include the updated information from recent developments
in hypernuclear physics. We take into account the two additional hidden-strangeness
mesons, $\sigma^{\ast}$ and $\phi$, and consider two cases of the $YY$ interaction.
The weak $YY$ interaction is suggested by the Nagara event, whereas the strong $YY$
interaction is deduced from the earlier measurement. By comparing the results with
different $YY$ interactions, we evaluate how sensitive the results
are to this factor.

\begin{figure}[htb]
\includegraphics[bb=20 330 480 770, width=8.6 cm,clip]{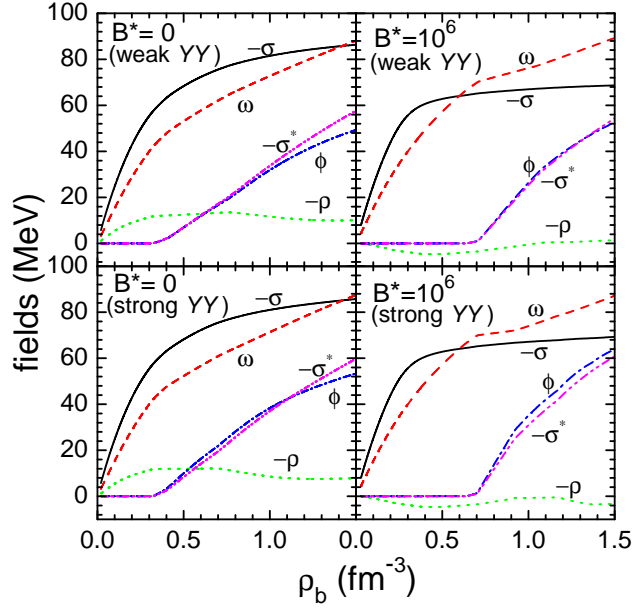}
\caption{(Color online) Meson mean fields as a function of
baryon density, for $B^{*}=10^{6}$ and the field-free ($B^{*}=0$) case,
with the weak and strong $YY$ interactions.}
\label{fig:field}
\end{figure}

\begin{figure}[htb]
\includegraphics[bb=20 330 480 770, width=8.6 cm,clip]{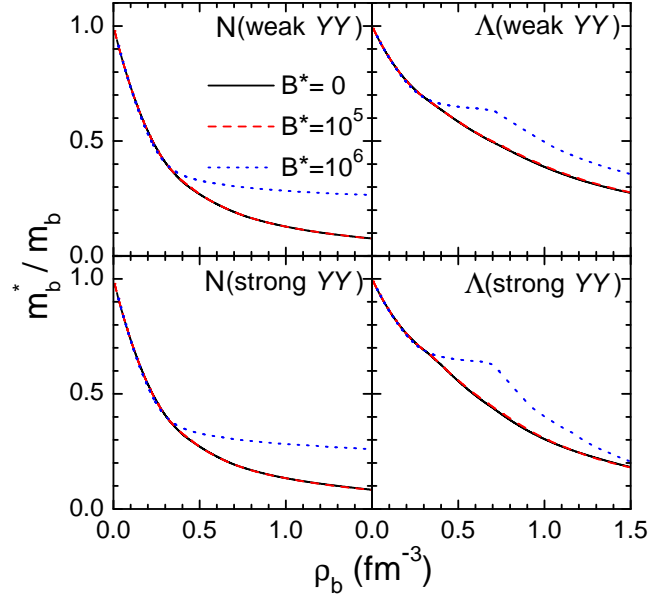}
\caption{(Color online) Effective masses of nucleons (N)
and $\Lambda$ hyperons as a function of baryon density
for $B^{\ast}=0$, $10^{5}$, and $10^{6}$, and for weak and
strong $YY$ interactions.}
\label{fig:rm}
\end{figure}

It has been found in Refs.~\cite{Latt00,Latt01,Mao03,CBP97,Mao06,YS06,Latt02} that
the properties of neutron star matter can be significantly affected only for
the magnetic field strength $B^{*}=B/B_{c}^{e} > 10^{5}$ ($B_{c}^{e}=4.414\times
10^{13}$ G is the electron critical field).
In Fig.~\ref{fig:field}, we plot the meson mean fields obtained as a function of
the baryon density. The results with $B^{*}=10^{6}$ are shown in the right panels,
whereas those in the field-free case are shown in the left panels.
We note that the mean fields of $\sigma$ and $\sigma^{\ast}$ mesons have negative
values in this calculation.
It is seen that $-\sigma$ in the right panels is smaller than the one in the left
panels, but $\omega$ is a little larger. $-\sigma^{\ast}$ and $\phi$ in the right
panels appear later than those in the left panels, because the onset of hyperons
shifts to higher density in the presence of strong magnetic fields.
By comparing the upper and lower panels, we find that $-\sigma^{\ast}$ and $\phi$
with the weak $YY$ interaction are smaller than those with the strong $YY$
interaction. This is because the hyperon couplings to $\sigma^{\ast}$
in the weak $YY$ case are smaller than those in the strong $YY$
case (see Table~\ref{tab:2}).
In Fig.~\ref{fig:rm}, we show the effective masses of nucleons (left panels)
and $\Lambda$ hyperons (right panels) as a function of the baryon density
for $B^{\ast}=0$, $10^{5}$, and $10^6$.
We notice that other hyperons ($\Sigma$ and $\Xi$) have a similar
behavior to $\Lambda$. It is found that the influence of the magnetic field
on the effective masses is not observable until $B^{*} > 10^{5}$.
The effective masses for $B^{*}=10^{6}$
are significantly larger than the field-free values. This is because smaller
$-\sigma$ and $-\sigma^{\ast}$ are obtained in the presence of strong magnetic
fields as shown in Fig.~\ref{fig:field}. Note that the effective mass of baryon
species $b$ is given by $m_{b}^{\ast}=m_{b}+g_{\sigma b}\sigma
+g_{\sigma^{\ast} b}\sigma^{\ast}$. The decrease of the effective mass with
increasing density is much larger in the left panels than in the right panels,
which is mainly due to $g_{\sigma N}>g_{\sigma\Lambda}$. Because the hyperon
couplings to $\sigma^{\ast}$ in the weak $YY$ case are smaller than
those in the strong $YY$ case (see Table~\ref{tab:2}), the effective masses
of $\Lambda$ in the upper right panel of Fig.~\ref{fig:rm} are larger than
those in the lower right panel. On the other hand, the effective masses of
nucleons are not affected by $\sigma^{\ast}$ due to $g_{\sigma^{\ast} N}=0$.

\begin{figure}[htb]
\includegraphics[bb=20 330 480 770, width=8.6 cm,clip]{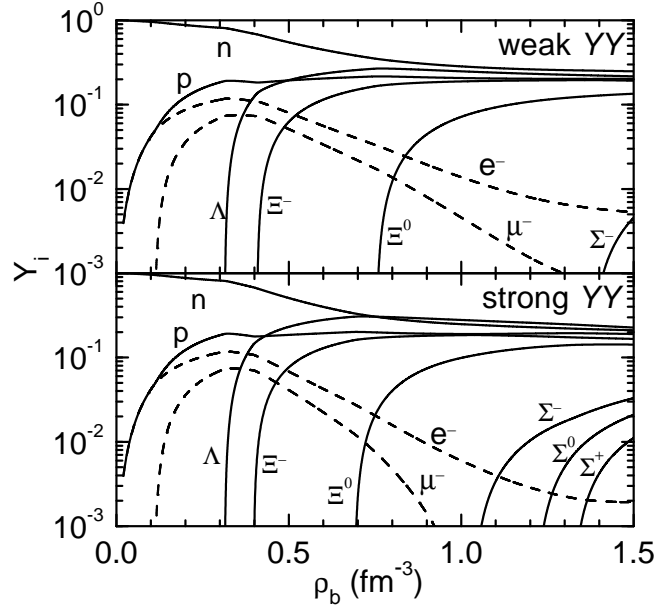}
\caption{Particle fraction $Y_{i}=\rho_{v}^{i}/\rho_{b}$
as a function of baryon density $\rho_{b}$ for $B^{*}=0$,
with weak and strong $YY$ interactions.}
\label{fig:Yi0}
\end{figure}

\begin{figure}[htb]
\includegraphics[bb=20 330 480 770, width=8.6 cm,clip]{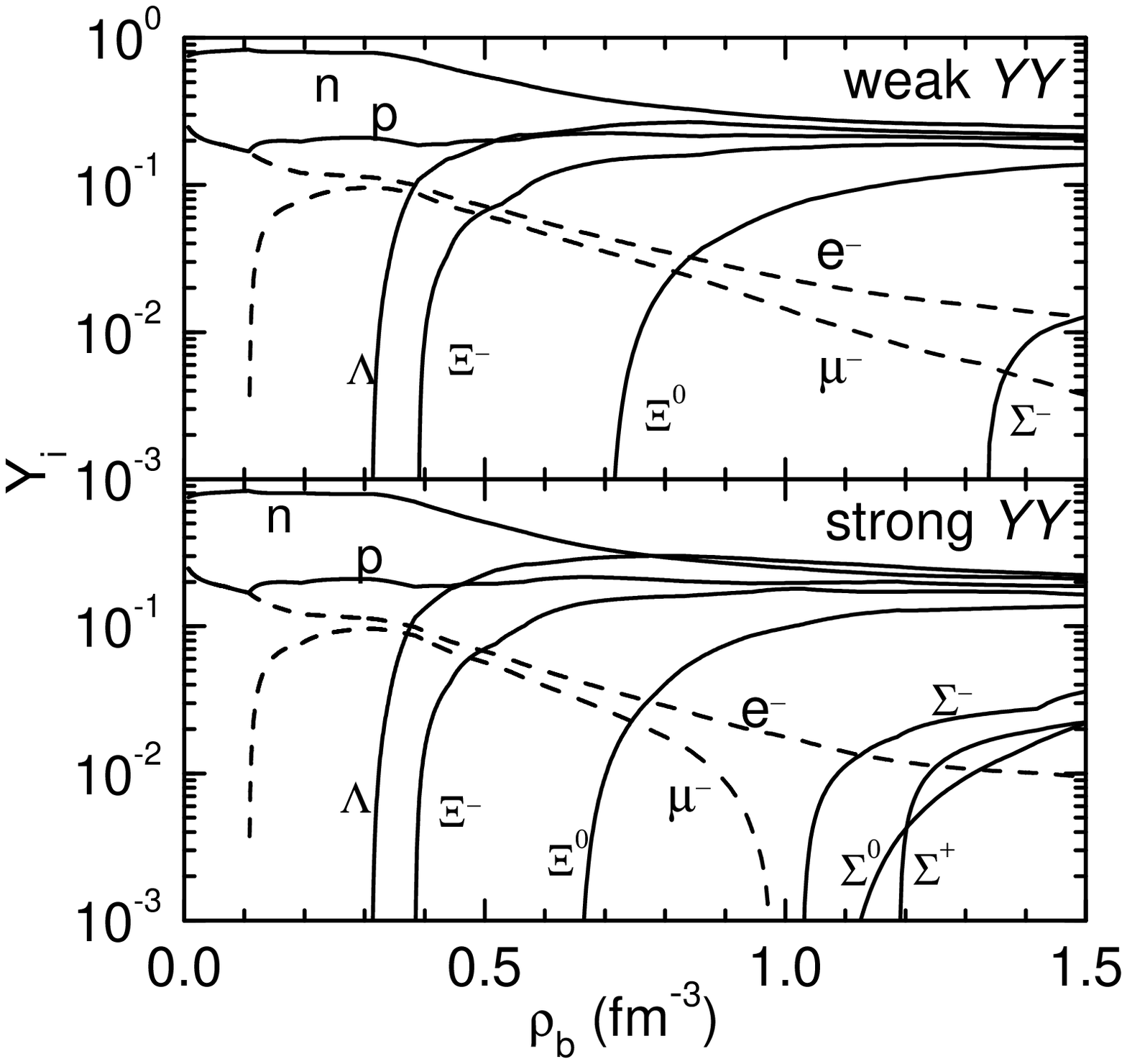}
\caption{Same as Fig.~\ref{fig:Yi0}, but for $B^{\ast}=10^{5}$.}
\label{fig:Yi5}
\end{figure}

\begin{figure}[htb]
\includegraphics[bb=20 330 480 770, width=8.6 cm,clip]{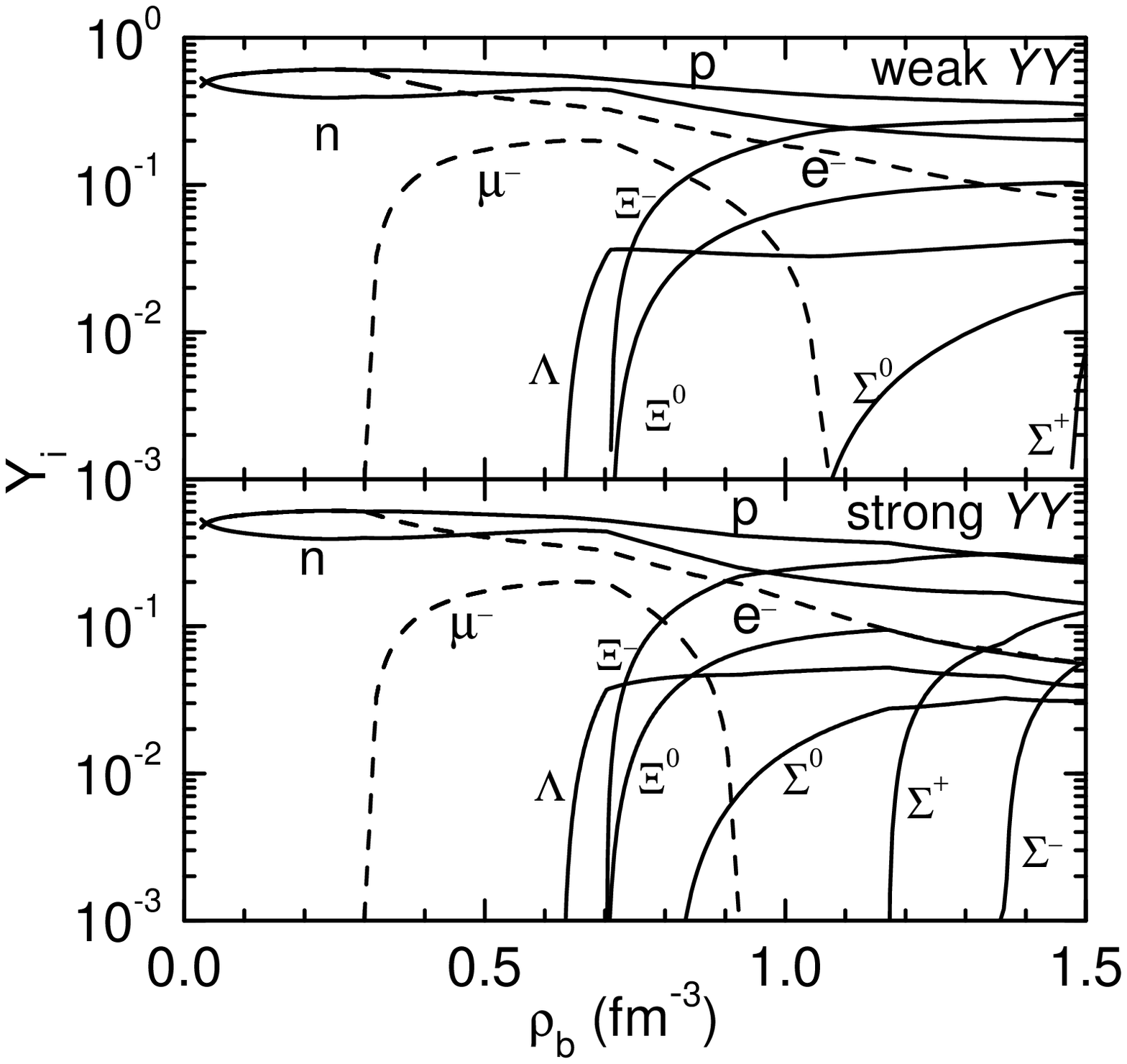}
\caption{Same as Fig.~\ref{fig:Yi0}, but for $B^{\ast}=10^{6}$.}
\label{fig:Yi6}
\end{figure}

In Figs.~\ref{fig:Yi0}, \ref{fig:Yi5}, and~\ref{fig:Yi6}, we present the particle
fraction $Y_{i}=\rho_{v}^{i}/\rho_{b}$ as a function of the baryon density
$\rho_{b}$ for $B^{*}=0$, $10^{5}$, and $10^{6}$, respectively.
At low densities, the fractions of nucleons and leptons are significantly
affected by the magnetic field. For $B^{*}=0$ shown in Fig.~\ref{fig:Yi0},
the proton fraction rises rapidly with increasing density and reaches $\sim$0.2
around $2\rho_{0}$ before the appearance of hyperons. For $B^{*}=10^{5}$
(see Fig.~\ref{fig:Yi5}), the proton fraction stays at $\sim$0.2
from low densities until the appearance of hyperons at $\sim$2$\rho_{0}$.
The proton fraction for $B^{*}=10^{6}$ (Fig.~\ref{fig:Yi6}) is larger than
$0.5$ before hyperons appear at $\sim$4$\rho_{0}$. It is obvious that
the threshold densities of hyperons are significantly altered
by the magnetic field in Fig.~\ref{fig:Yi6}. $\Lambda$ is the first hyperon
to appear in all cases. In Figs.~\ref{fig:Yi0} and~\ref{fig:Yi5}, $Y_{\Lambda}$
is found to be larger than other hyperon fractions, but it is much smaller
than $Y_{\Xi^{-}}$ and $Y_{\Xi^{0}}$ at higher densities in Fig.~\ref{fig:Yi6}.
This is because $\Lambda$ is a neutral particle and has a small anomalous
magnetic moment. In general, charged particles depend more on the magnetic
field than neutral particles due to the Landau quantization of charged particles.
For neutral particles, the anomalous magnetic moment plays an important role
in the presence of strong magnetic fields. We find that hyperon fractions are
very sensitive to hyperon couplings used in the calculation.
The appearance of hyperons leads to pronounced changes in the fractions of
nucleons and leptons. The negatively charged hyperons can play the same role
as electrons and muons in maintaining the electric charge neutrality.
Therefore, the appearance of $\Xi^{-}$ decreases $Y_{e}$ and $Y_{\mu}$.
By comparing the upper and lower panels of Figs.~\ref{fig:Yi0}--\ref{fig:Yi6},
we find that the differences between weak and strong $YY$
cases are not very large and mainly exist at high densities.

\begin{figure}[htb]
\includegraphics[bb=10 330 480 770, width=8.6 cm,clip]{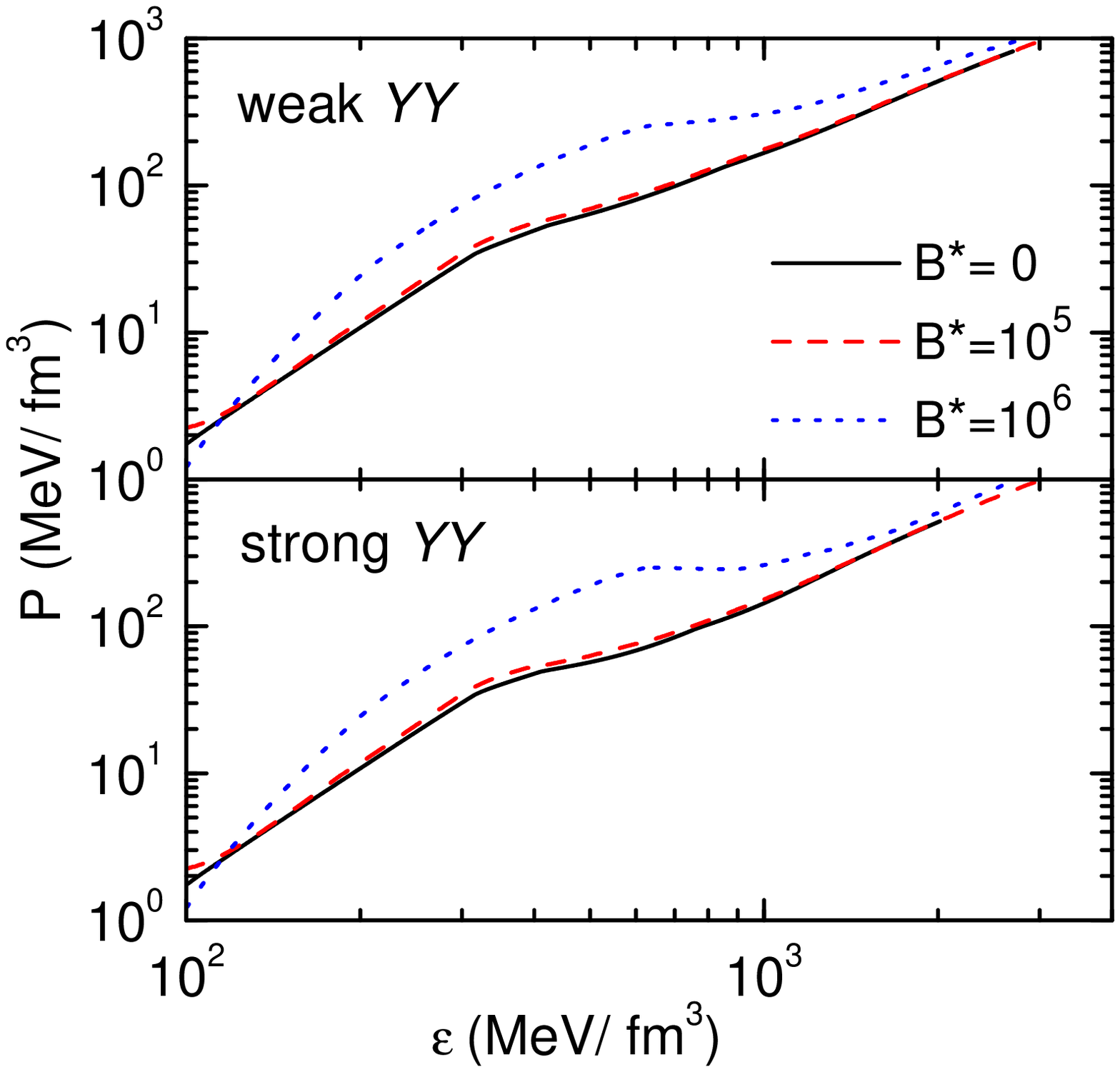}
\caption{(Color online) Matter pressure $P$ vs the matter
energy density $\varepsilon$ for $B^{\ast}=0$, $10^{5}$, and $10^{6}$,
for weak and strong $YY$ interactions.}
\label{fig:ep}
\end{figure}

In Fig.~\ref{fig:ep}, we show the matter pressure $P$ as a function of the
matter energy density ${\varepsilon}$ for the magnetic field strengths
$B^{*}=0$, $10^{5}$, and $10^{6}$. The results with the weak and strong $YY$
interactions are plotted in the upper and lower panels, respectively.
It is found that the EOS with the weak $YY$ interaction is slightly stiffer
than the one with the strong $YY$ interaction. The presence of hyperons makes
the EOS softer than in the case without hyperons, which has been discussed
in our previous work~\cite{YS06}. The softening of the EOS becomes less
pronounced with increasing magnetic field for $B^{*}>10^{5}$. This is because
the onset of hyperons shifts to higher densities, and the hyperon contribution
becomes smaller with increasing $B^{*}$. Here we include the anomalous magnetic
moments of all baryons, which play an important role in determining
the EOS and composition of hyperonic matter.


\section{Summary}

\label{sec:5}

In this article, we have studied the effects of strong magnetic fields on the
properties of hyperonic matter. We have employed the
RMF theory with the TM1 parameter set which is known to provide excellent
descriptions of nuclear matter and finite nuclei including unstable nuclei.
In the RMF approach, baryons interact through the exchange
of scalar and vector mesons.
The baryons considered in this work are nucleons ($p$ and $n$) and hyperons
($\Lambda$, $\Sigma$, and $\Xi$). The exchanged mesons include isoscalar scalar
and vector mesons ($\sigma$ and $\omega$), isovector vector meson ($\rho$),
and two additional hidden-strangeness mesons ($\sigma^{\ast}$ and $\phi$).
It is well known that the meson-hyperon couplings play an important role
in determining the properties of hyperonic matter. We have used the couplings
constrained by reasonable hyperon potentials that include the updated information
from recent developments in hypernuclear physics. The weak hyperon-hyperon ($YY$)
interaction suggested by the Nagara event has been adopted to investigate the
properties of hyperonic matter, whereas the strong $YY$ interaction deduced
from the earlier measurement has also been used for comparison. We found that
the EOS with the weak $YY$ interaction is slightly stiffer than the one with
the strong $YY$ interaction, and the differences mainly exist at high densities.

It is found that the effects of strong magnetic fields become significant
only for magnetic field strength $B^{*}>10^{5}$. The threshold densities of
hyperons can be significantly altered by the magnetic field.
The presence of hyperons makes the EOS softer than in the case without
hyperons, and the softening of the EOS becomes less pronounced with increasing
$B^{*}$. We found that $\Lambda$ is the first hyperon to appear in all cases
considered here. For strong magnetic fields as shown in Fig.~\ref{fig:Yi6},
the fractions of $\Lambda$ are much smaller than those of $\Xi^{-}$ and $\Xi^{0}$
at higher densities. This is because $\Lambda$ is a neutral particle and has a small
anomalous magnetic moment. It is found that charged particles depend more on the
magnetic field than neutral particles due to the Landau quantization of charged
particles. In this work, we have included the anomalous magnetic moments
of all baryons, which play an important role in determining the EOS and
composition of hyperonic matter.

\section*{Acknowledgment}

This work was supported in part by the National Natural Science
Foundation of China (No. 10675064).


\newpage
\begin{table}[tbp]
\caption{Static properties of baryons considered in this work.
The mass and charge of baryon species $b$ are denoted by $m_b$
and $q_b$, respectively. The baryonic magnetic moment is denoted
by $\mu_b$, and the anomalous magnetic moment is given
by $\kappa_b = (\mu_b/\mu_N - q_b m_p/m_b)\mu_N$, where $\mu_N=e\hbar/2m_p$
is the nuclear magneton.}
\begin{center}
\begin{tabular}{lrrrr}
\hline\hline
$b$        & $m_b$    & $q_b$& $\mu_b$ & $\kappa_b$ \vspace*{-0.2cm} \\
           & (MeV)    & ($e$)&($\mu_N$)&($\mu_N$)   \\
\hline
$p$        &  $938.0$ &  $1$ & $2.79$ & $1.79$ \\
$n$        &  $938.0$ &  $0$ &$-1.91$ &$-1.91$ \\
$\Lambda$  & $1115.7$ &  $0$ &$-0.61$ &$-0.61$ \\
$\Sigma^+$ & $1193.1$ &  $1$ & $2.46$ & $1.67$ \\
$\Sigma^0$ & $1193.1$ &  $0$ & $1.61$ & $1.61$ \\
$\Sigma^-$ & $1193.1$ & $-1$ &$-1.16$ &$-0.37$ \\
$\Xi^0$    & $1318.1$ &  $0$ &$-1.25$ &$-1.25$ \\
$\Xi^-$    & $1318.1$ & $-1$ &$-0.65$ & $0.06$ \\
\hline\hline
\end{tabular}
\label{tab:1}
\end{center}
\end{table}

\begin{table}[tbp]
\caption{Scalar coupling constants determined by the hyperon potentials.
We take $g_{\sigma^*\Sigma}=g_{\sigma^*\Lambda}$ in this calculation. }
\begin{center}
\begin{tabular}{cccccc}
\hline\hline
      &$g_{\sigma\Lambda}$ &$g_{\sigma\Sigma}$ &$g_{\sigma\Xi}$ &
      $g_{\sigma^*\Lambda}$ & $g_{\sigma^*\Xi}$ \\
\hline
   Weak $YY$   & 6.228 & 4.472 & 3.114 & 5.499 & 11.655  \\
 Strong $YY$   & 6.228 & 4.472 & 3.114 & 7.103 & 12.737  \\
\hline\hline
\end{tabular}
\label{tab:2}
\end{center}
\end{table}

\newpage

\begin{thebibliography}{99}
\bibitem{Weber05} F. Weber, Prog. Part. Nucl. Phys. \textbf{54}, 193 (2005).

\bibitem{HL06} A. K. Harding and D. Lai, Rep. Prog. Phys. \textbf{69}, 2631 (2006).

\bibitem{TD96} C. Thompson and R. C. Duncan, Astrophys. J. \textbf{473}, 322 (1996).

\bibitem{Latt00} A. Broderick, M. Prakash, and J. M. Lattimer,
Astrophys. J. \textbf{537}, 351 (2000).

\bibitem{Latt01} C. Y. Cardall, M. Prakash, and J. M. Lattimer,
Astrophys. J. \textbf{554}, 322 (2001).

\bibitem{Mao03} G. J. Mao, A. Iwamoto, and Z. X. Li,
Chin. J. Astron. Astrophys. \textbf{3}, 359 (2003).

\bibitem{CBP97} S. Chakrabarty, D. Bandyopadhyay, and S. Pal,
Phys. Rev. Lett. \textbf{78}, 2898 (1997).

\bibitem{Mao06} F. X. Wei, G. J. Mao, C. M. Ko, L. S. Kisslinger, H. St\"{o}cker,
and W. Greiner, J. Phys. G \textbf{32}, 47 (2006).

\bibitem{YS06} P. Yue and H. Shen, Phys. Rev. C \textbf{74}, 045807 (2006).

\bibitem{Latt02} A. Broderick, M. Prakash, and J. M. Lattimer,
Phys. Lett. \textbf{B531}, 167 (2002).

\bibitem{Pion01} I.-S. Suh and G. J. Mathews, Astrophys. J. \textbf{546}, 1126 (2001).

\bibitem{Pion07} K. Takahashi, J. Phys. G \textbf{34}, 653 (2007).

\bibitem{Kaon02} P. Dey, A. Bhattacharyya, and D. Bandyopadhyay,
J. Phys. G \textbf{28}, 2179 (2002).

\bibitem{YS08} P. Yue and H. Shen, Phys. Rev. C \textbf{77}, 045804 (2008).

\bibitem{CSCPRL05NPB06} E. J. Ferrer, V. de la Incera, and C. Manuel,
Phys. Rev. Lett. \textbf{95}, 152002 (2005); Nucl. Phys. \textbf{B747}, 88 (2006).

\bibitem{CSCPRL06PRD07} E. J. Ferrer and V. de la Incera,
Phys. Rev. Lett. \textbf{97}, 122301 (2006);
Phys. Rev. D \textbf{76}, 045011 (2007); Phys. Rev. D \textbf{76}, 114012 (2007).

\bibitem{CSCPRD07} J. L. Noronha and I. A. Shovkovy,
Phys. Rev. D \textbf{76}, 105030 (2007).

\bibitem{CSCPRL08} K. Fukushima and H. J. Warringa,
Phys. Rev. Lett. \textbf{100}, 032007 (2008).

\bibitem{PR00} H. Heiselberg and M. Hjorth-Jensen,
Phys. Rep. \textbf{328}, 237 (2000).

\bibitem{PRC96} J. Schaffner and I. N. Mishustin,
Phys. Rev. C \textbf{53}, 1416 (1996).

\bibitem{PRC99} S. Pal, M. Hanauske, I. Zakout, H. St\"{o}cker, and W. Greiner,
Phys. Rev. C \textbf{60}, 015802 (1999).

\bibitem{Panda04} P. K. Panda, D. P. Menezes, and C. Providencia,
Phys. Rev. C \textbf{69}, 025207 (2004).

\bibitem{Shen02} H. Shen, Phys. Rev. C \textbf{65}, 035802 (2002).

\bibitem{Gl01} N. K. Glendenning, Phys. Rev. C \textbf{64}, 025801 (2001).

\bibitem{Shen03} H. Shen and Z. L. Zhang, Chin. Phys. Lett. \textbf{20}, 650 (2003).

\bibitem{KN86} D. B. Kaplan and A. E. Nelson, Phys. Lett. \textbf{B175}, 57 (1986).

\bibitem{PRC08} F. Yang and H. Shen, Phys. Rev. C \textbf{77}, 025801 (2008).

\bibitem{Nagara} H. Takahashi \textit{et al}.,
Phys. Rev. Lett. \textbf{87}, 212502 (2001).

\bibitem{PRC03s} H. Q. Song, R. K. Su, D. H. Lu, and W. L. Qian,
Phys. Rev. C \textbf{68}, 055201 (2003).

\bibitem{JPG04s} W. L. Qian, R. K. Su, and H. Q. Song,
J. Phys. G \textbf{30}, 1893 (2004).

\bibitem{JPG05} I. Bednarek and R. Manka, J. Phys. G \textbf{31}, 1009 (2005).

\bibitem{Serot86} B. D. Serot and J. D. Walecka,
Adv. Nucl. Phys. \textbf{16}, 1 (1986).

\bibitem{Ring90} Y. K. Gambhir, P. Ring, and A. Thimet,
Ann. Phys. (N.Y.) \textbf{198}, 132 (1990).

\bibitem{Toki96} D. Hirata, K. Sumiyoshi, B. V. Carlson, H. Toki, and I. Tanihata,
Nucl. Phys. \textbf{A609}, 131 (1996).

\bibitem{Ren02}  Z. Z. Ren, F. Tai, and D. H. Chen,
Phys. Rev. C \textbf{66}, 064306 (2002).

\bibitem{Shen06} H. Shen, F. Yang, and H. Toki,
Prog. Theor. Phys. \textbf{115}, 325 (2006).

\bibitem{Shen98} H. Shen, H. Toki, K. Oyamatsu, and K. Sumiyoshi,
Nucl. Phys. \textbf{A637}, 435 (1998).

\bibitem{PRL93} J. Schaffner, C. B. Dover, A. Gal, C. Greiner, and H. St\"{o}cker,
Phys. Rev. Lett. \textbf{71}, 1328 (1993).

\bibitem{ST94} Y. Sugahara and H. Toki, Nucl. Phys. \textbf{A579}, 557 (1994).

\bibitem{HTT95} D. Hirata, H. Toki, and I. Tanihata,
Nucl. Phys. {\bf A589}, 239 (1995).

\bibitem{MT97} Z. Y. Ma,  H. Toki, B. Q. Chen, and N. Van Giai,
Prog. Theor. Phys. {\bf 98}, 917 (1997).

\bibitem{MV01} Z. Y. Ma,  N. Van Giai, A. Wandelt, D. Vretenar, and P. Ring,
Nucl. Phys. \textbf{A686}, 173 (2001).

\bibitem{eAMM} R. C. Duncan, arXiv:astro-ph/0002442.

\bibitem{PRD71} W. Y. Tsai and A. Yildiz, Phys. Rev. D \textbf{4}, 3643 (1971).

\bibitem{PRC00} J. Schaffner-Bielich and A. Gal,
Phys. Rev. C \textbf{62}, 034311 (2000).

\bibitem{JPG08} C. Ishizuka, A. Ohnishi, K. Tsubakihara, K. Sumiyoshi, and S. Yamada,
J. Phys. G \textbf{35}, 085201 (2008).

\bibitem{PRC00xi} P. Khaustov \textit{et al}.,
Phys. Rev. C \textbf{61}, 054603 (2000).

\bibitem{ANN94} J. Schaffner, C. B. Dover, A. Gal, C. Greiner, D. J.
Millener, and H. St\"{o}cker, Ann. Phys. (N.Y.) \textbf{235}, 35 (1994).

\end{thebibliography}
\end{document}